\documentclass[preprint,1p]{elsarticle}
\usepackage{amssymb,graphicx}

\newcommand*{\units}[1]{\,\mbox{#1}}
\newcommand*{\half}{\frac{1}{2}}

\newcommand*{\ket}[1]{| #1 \rangle}

\newcommand*{\ketbra}[2]{| #1 \rangle\langle #2 |}

\biboptions{sort&compress}

\begin{document}
\begin{frontmatter}

\title{Cavity Quantum Electrodynamics with Quantum Interference in a Three-level 
Atomic System}

\author{Amitabh Joshi\corref{cor1}}
\ead{mcbamji@gmail.com}
\address{Department of Physics and Optical Engineering, Rose-Hulman Institute of 
Technology, Terre Haute, Indiana 47803, USA}

\author{Juan D. Serna\corref{cor2}}
\ead{juan.serna@scranton.edu}
\address{Department of Physics and Electrical Engineering, University of 
Scranton, Scranton, Pennsylvania 18510, USA}

\cortext[cor1]{Principal corresponding author}
\cortext[cor2]{Corresponding author}

\begin{abstract}
Spontaneously generated coherence and enhanced dispersion in a V-type, 
three-level atomic system interacting with a single mode field can considerably 
reduce the radiative and cavity decay rates. This may eliminate the use of high 
finesse, miniaturized cavities in optical cavity quantum electrodynamics 
experiments under strong atom-field coupling conditions.
\end{abstract}

\end{frontmatter}

\section{\label{sec:intro}Introduction}
In experiments related to cavity quantum electrodynamics (QED), it is required 
to deal with three important parameters governing the atom-field dynamics, which 
are the atom-field coupling strength $g$, the radiative decay rate of the atom 
$\gamma_{a}$, and the cavity decay rate $\gamma_{cav}$. The parameter $g$ 
characterizes the oscillatory exchange of excitation between the atom and the 
cavity field mode while its magnitude relative to the parameters $\gamma_a$ and 
$\gamma_{cav}$ decides if the coupling between atom and resonator is weak or 
strong. Both the suppression and enhancement of radiative decay rate $\gamma_a$ 
have been observed in cavities subtending a very large solid angle over the atom 
under the weak coupling regime $\gamma_{a} \ll g^{2} / \gamma_{cav}\ll 
\gamma_{cav}$, which is in good agreement with the perturbation 
theory~\cite{Purcell:1946.69,Kleppner:1981.47,Haroche:1985.20,Haroche:1989.42}. 
The strong coupling regime has also been successfully realized with Rydberg 
atoms in the microwave domain where the condition $g^{2}/\gamma_{cav} \gg 
\gamma_{cav} > \gamma_{a}$ is readily met and the Rydberg atoms possess very 
large dipole moments, long radiative decay times (using circular Rydberg 
states), and a superconducting cavity operating at 10$^{0}$K is 
employed~\cite{Haroche:1992}. For cavity QED experiments in the optical domain, 
the strong coupling regime is usually very difficult to achieve since the 
coupling parameter $g$ is intrinsically weak in comparison to $\gamma_a$ and 
$\gamma_{cav}$. One can express the coupling parameter as $g=(\mu^{2}\omega_{0} 
/ 2\hbar\epsilon_{0}V)^{1/2}$, where $\mu$ is the transition dipole moment and 
$V$ is the effective cavity mode volume. Since the magnitude of $g$ is small in 
the optical domain, one needs to greatly reduce $V$ to achieve the strong 
coupling regime~\cite{Berman:1994,Vernooy:1999.83}. Alternatively, one can 
define the critical photon number $m_0=\gamma_a ^{2}/2g^{2}$ and the critical 
atom number $N_0=2\gamma_a \gamma_{cav}/g^{2}$ to do nonlinear optics with one 
photon per mode and single-atom switching for optical cavity response. In these 
cases, the parameter $g$ (internal interaction strength) is responsible for the 
information exchange while $\gamma_a$ and $\gamma_{cav}$ (external 
dephasing/dissipative rates) are responsible for the rate of information loss 
from the system. For strong coupling regimes, it is required to have $m_0 \ll 1$ 
and $N_0 \ll 1$. This implies that the mode volume should be as small as 
possible and the photon leakage rate ($\gamma_{cav}$) should also be small, 
meaning that a very high-Q cavity with very large finesse is required. Various 
scaling configurations such as the \emph{hourglass cavity} has been employed for 
this purpose. The record finesse of $3\times 10^{6}$ has been achieved with 
$m_0=8\times 10^{-6}$ and $N_0=7\times 10^{-4}$; and $g=110\units{MHz}$, 
$\gamma_a=2.6\units{MHz}$, $\gamma_{cav} =14.2\units{MHz}$ are reported for the 
cavity QED experiments with Cs atomic beams~\cite{Hood:2000.287,Hood:2001.64}.

In this work, we propose an alternative way to reach the strong coupling regime 
in the optical domain for cavity QED experiments using the spontaneously 
generated coherence (SGC) in a three-level atom (V-type) to reduce the radiative 
decay rate ($\gamma_{a}$)~\cite{Cardimona:1989.22,Zhou:1996.77}. The reduction 
of the cavity decay rate ($\gamma_{cav}$) comes due to the large dispersion 
(which can easily exceed the empty cavity dispersion in the case of optically 
thick medium) near to the point of almost-vanishing 
absorption~\cite{Lukin:1998.23,Wang:2000.25,Goorskey:2002.49}. The system can 
quench the fluorescence for all frequencies under the condition of maximum 
quantum interference if the detuning satisfies certain condition. Due to the 
quantum interference (i.e., SGC), we observe a rapid change in the refractive 
index near the vanishing absorption. Here the medium provides a large dispersion 
capable of reducing the cavity linewidth. Because of this, we can achieve the 
strong coupling conditions not by decreasing the effective mode volume, but by 
reducing the atomic decay rate ($\gamma_a$) via quantum interference, and cavity 
decay rate ($\gamma_{cav}$) through enhanced dispersion. Both of these 
properties have been studied recently under the induced atomic coherence and 
quantum interference in three-level atomic 
systems~\cite{Cardimona:1989.22,Zhou:1996.77,Lukin:1998.23,Wang:2000.25,
Goorskey:2002.49}. With this proposed scheme, one can investigate interesting 
cavity-QED effects in the optical domain with sizeable optical cavities 
containing atomic cells, that can be used to manipulate photon and atomic states 
for quantum information processing.

Recently, a hybrid absorptive-dispersive, atomic optical bistability in an open 
$\Lambda$-type, three-level system was studied using a microwave field to drive 
the hyperfine transition between two lower states, and including the incoherent 
pumping and spontaneously generated coherence~\cite{Wang:2012.29}. In another 
work, the resonance fluorescence spectrum of a three-level, ladder system driven 
by two laser fields was investigated and its resemblance with a V-type system 
with parallel dipole moments was compared. The ladder system was experimentally 
studied using a $^{85}$Rb atom beam, which showed the narrowing of the central 
peak and reminding the spontaneously generated coherence phenomenon in a V-type 
system responsible for such narrowing~\cite{Tian:2012.21}. Likewise, a 
two-mode-entangled light generation from a laser-driven, three-level V-type atom 
kept inside a cavity was reported, where the spontaneously generated quantum 
interference between two atomic decay channels played a crucial 
role~\cite{Tan:2009.79}. In reality, there is continued interest to generate SGC 
in cavity QED. For example, SGC was experimentally observed via its effect on 
the absorption spectrum in a rubidium atomic beam without imposing the rigorous 
requirement of close-lying levels. The experiments were carried out both in a 
four-level, N-type and four-level, inverted-Y-type rubidium atomic 
systems~\cite{Tian:2012.285}. In a recent work, generation of SGC in a Rb atomic 
system was proposed using photon counting statistics in a four-level, Y-type 
model driven by three coherent fields; ultra narrow probe absorption peaks in 
the presence of SGC were also predicted~\cite{Song:2016.015001}. The rest of 
the paper is organized as follows. In section~\ref{sec:model}, we discuss the 
model, equations proposed, and results. Some concluding remarks are given in 
section~\ref{sec:conclusion}.

\section{\label{sec:model}Model, equations, and results}
\begin{figure}
  \centering
  \includegraphics[scale=0.5]{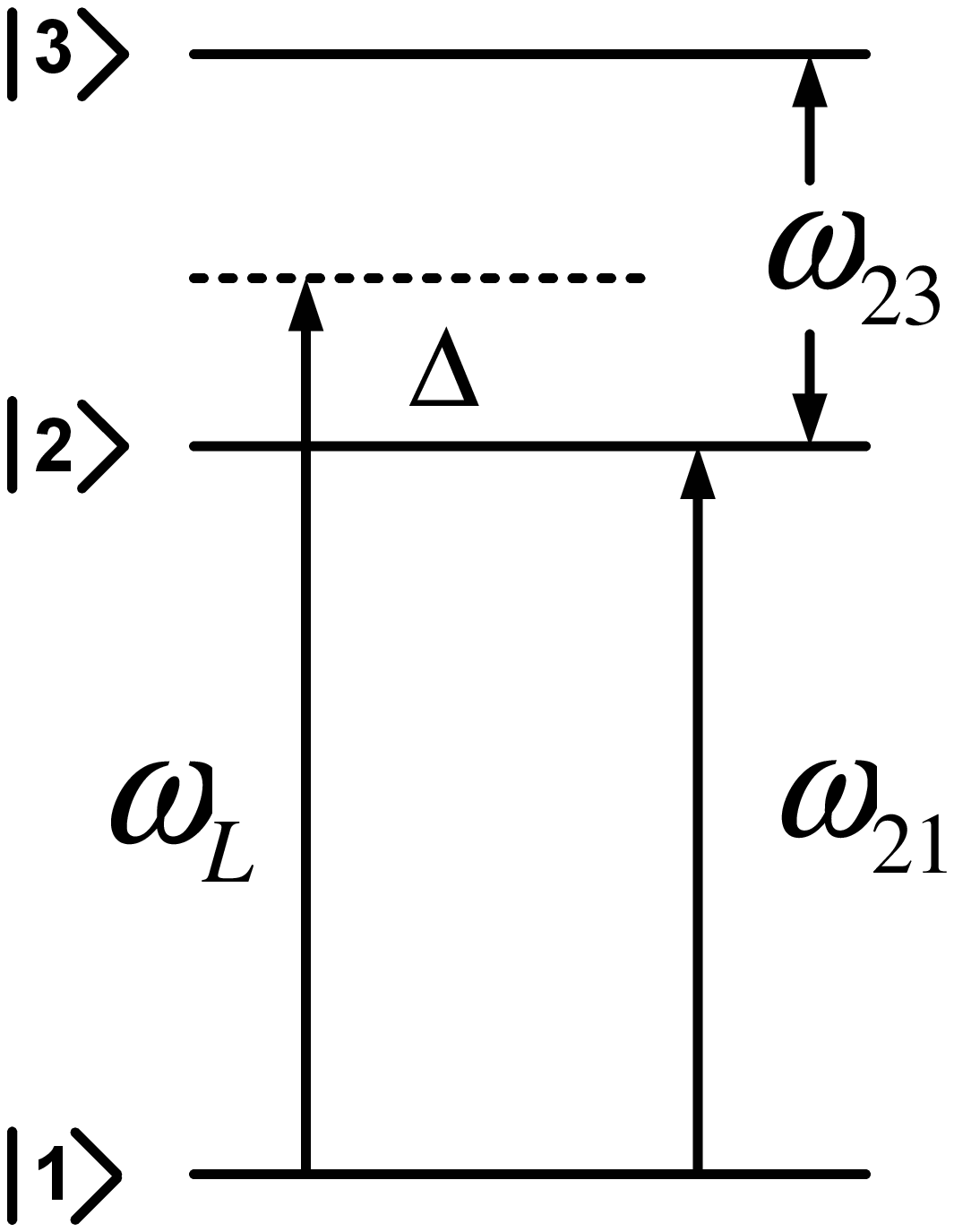}
  \caption{Diagram of a three-level system in V-configuration driven by a laser
  of frequency $\omega _{L}$.}
  \label{fig:Fig01}
\end{figure}

In this work, we consider a model atom in a V-type configuration of its levels, 
consisting of two upper states $\ket{2}$ and $\ket{3}$, coupled to a common 
lower level $\ket{1}$ by a single-mode laser field with amplitude $E_{L}$ and 
frequency $\omega _{L}$ (see Fig.~\ref{fig:Fig01}). The Hamiltonian of the 
system in the rotating frame of the field of frequency $\omega _{L}$ is given by
\begin{equation}
H=(\Delta -\omega _{23}) \ketbra{2}{2} + \Delta \ketbra{3}{3} +
  \left[\left(\Omega_{1} \ketbra{2}{1} + \Omega _{2} \ketbra{3}{1}\right) + 
   H.c.\right],
\end{equation}
where $\Delta =\omega_{21}-\omega_{L}$ is the detuning of the laser field 
frequency from level $\ket{2}$, $\Omega_{k}=2(d_{k+1,1})E_{L}/\hbar$ ($k=1,2$) 
is the Rabi frequency, $d_{k+1,1}$ is the dipole matrix element of the atomic 
transition from $\ket{1}$ to $\ket{k+1}$ ($k=1,2$), and $\omega_{23}$ is the 
level splitting of the upper levels. $\ketbra{m}{n}$ is the dipole transition 
operator when $m \neq n$ (or the population operator when $m=n$). In the frame 
rotating with the applied field, the equations of motion of the reduced density 
matrix elements are
\begin{eqnarray}
\dot{\rho}_{11} &=& 2\gamma_{1} \rho_{22} + 2\gamma_{2} \rho_{33} +
2\gamma_{12}(\rho_{23}+\rho_{32}) + i \frac{\Omega_{1}}{2}(\rho_{12} -
\rho_{21}) + i\frac{\Omega_{2}}{2}(\rho_{13}-\rho_{31}), \nonumber\\
\dot{\rho}_{22} &=& -2\gamma_{1}\rho_{22} - \gamma_{12}(\rho_{23} +
\rho_{32}) - i\frac{\Omega_{1}}{2}(\rho_{12}-\rho_{21}), \nonumber\\
\dot{\rho}_{33} &=& -2\gamma_{2}\rho_{33} - \gamma_{12}(\rho_{23}+
\rho_{32}) - i\frac{\Omega_{2}}{2}(\rho_{13}-\rho_{31}), \nonumber\\
\dot{\rho}_{21} &=& -(i\Delta + \gamma_{1})\rho_{21} + 
i\frac{\Omega_{1}}{2}(\rho_{22} -\rho_{11})+ i\frac{\Omega_{2}}{2}\rho_{23} 
-\gamma_{12}\rho_{31}, \nonumber\\
\dot{\rho}_{32} &=& -(-i\omega_{23} + \gamma_{1}+\gamma_{2})\rho_{32} + 
i\frac{\Omega_{1}}{2}\rho_{31} - i\frac{\Omega_{2}}{2}\rho_{12} -\gamma_{12} 
(\rho_{22}+\rho_{33}), \nonumber\\
\dot{\rho}_{31} &=& -(i(\Delta -\omega_{23})+ \gamma_{2}) \rho_{31} +
i\frac{\Omega_{2}}{2}(\rho_{33} -\rho_{11})+ i\frac{\Omega_{1}}{2}
\rho_{32} -\gamma_{12}\rho_{21},
\label{eqn:rDensity}
\end{eqnarray}
in which $\gamma_{k}$ is the spontaneous decay constant of the excited upper 
levels $k+1$ ($k=1,2$) to the ground level $\ket{1}$. The term $\gamma_{12}$ 
accounts for the spontaneous emission induced quantum interference effect due to 
the cross coupling between emission processes in the radiative channels $\ket{2} 
\rightarrow \ket{1}$ and $\ket{3} \rightarrow \ket{1}$. The quantum interference 
terms in (\ref{eqn:rDensity}) represent the physical situation in which a photon 
is emitted virtually in channel $\ket{2} \rightarrow \ket{1}$ and virtually 
absorbed in channel $\ket{1} \rightarrow \ket{3},$ or vice versa. 
Equation~(\ref{eqn:rDensity}) can be written in the Lindblad form. The details 
of such equation are mentioned in Ref~\cite{Tang:2010.82}.

Quantum interference plays a very significant role in spectral line narrowing, 
fluorescence quenching, population trapping, etc. Although in a recent 
experiment the ability of controlling $\gamma_{12}$ has been experimentally 
demonstrated in sodium dimers by considering the superposition of singlet and 
triplet states due to spin-orbit coupling~\cite{Xia:1996.77},  a conflicting 
result was obtained in another experiment of similar kind~\cite{Li:2000.84}. 
However, SGC was observed in an absorption experiment using rubidium atomic 
beam~\cite{Tian:2012.285}. The quantum interference effect is sensitive to the 
atomic dipole orientation. If dipoles $\vec d_{21}$ and $\vec d_{31}$ are 
parallel to each other, then $\gamma_{12}= \sqrt {\gamma_{1}\gamma_{2}}$, and 
the interference is maximal. On the other hand, if $\vec d_{21}$ and $\vec 
d_{31}$ are perpendicular to each other, then $\gamma_{12}=0$, and there is no 
quantum interference. We can see this more clearly by exploring the origin of 
such coherence. The photon emitted during spontaneous emission on one of the two 
atomic transitions in the system drives the other transition. The strength 
parameter of the coherence, represented by the coefficient $\gamma_{12}$, is 
directly proportional to the mutual polarization of the transition dipole 
moments of the two transitions characterized by $p = \cos\theta$, where $\theta$ 
is the angle between the two dipole moments. One can write $\gamma_{12} = \sqrt 
{\gamma_1 \gamma_2} \cos\theta$. If the two transition dipole moments are 
perpendicular to each other, then $p=0$ and $\gamma_{12}=0$, leading to no SGC. 
Similarly, if the dipole moments are parallel to each other, then $p = 1$ and 
$\gamma_{12}=1$, leading to maximum SGC~\cite{Tang:2010.82}.
In addition to 
this, one can also have partial quantum interference.

\begin{figure}
  \centering
  \includegraphics{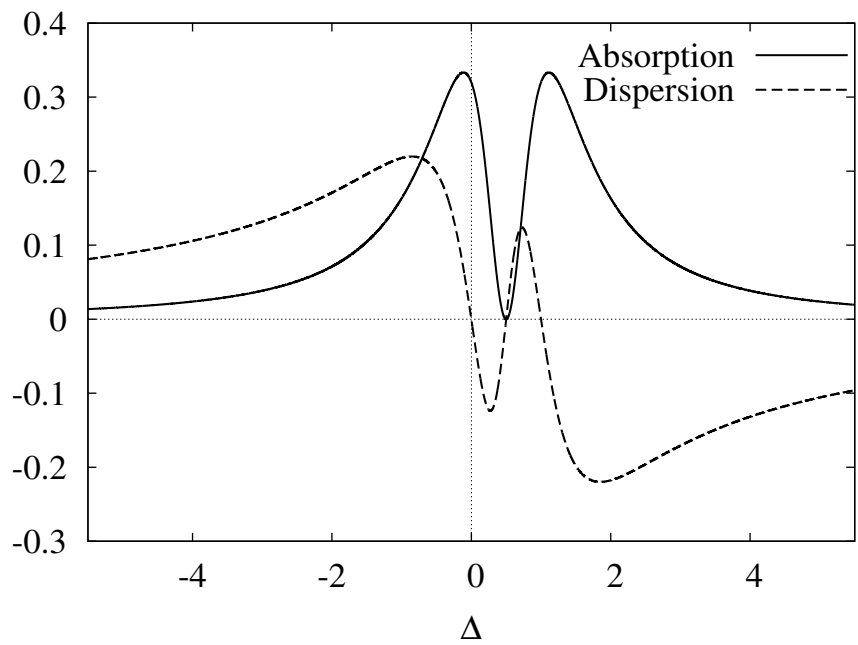}
  \caption{Absorption $A$ and dispersion $\eta$ versus detuning $\Delta$ as 
   given by equations (\ref{eq:absoption},~\ref{eq:dispersion}) for $\gamma = 
   0.5$, $\Omega = 0.5$, and $\omega_{23} = 1$. The left and right peaks for 
   the absorption curve occur at $\Delta = -0.1,\,1.1$, respectively. The
   trough is located at $\Delta = 0.5$.}
  \label{fig:Fig02}
\end{figure}

The absorption (dispersion) spectrum is proportional to the real (imaginary) 
part of the term $\rho _{21}+\rho _{31}$. Using (\ref{eqn:rDensity}), it is 
straightforward to evaluate analytically this term in the steady state assuming 
that the dipole moments are equal and parallel. Assuming that $\Omega_{1}$ and 
$\Omega_{2}$ are real and equal ($\Omega_{1} = \Omega_{2} = \Omega$), and 
setting the two radiative damping constants to be equal, $\gamma_{1} = 
\gamma_{2} = \gamma$, the steady-state absorption and dispersion are found to be
\begin{equation}
\label{eq:absoption}
A(\Delta) = \frac{2\gamma\Omega(\omega_{23}-2\Delta)^{2}}
{4\Delta^{2}(\omega_{23} - \Delta)^{2} + 2\Omega^{2}(\omega_{23}^{2}-
2\omega_{23}\Delta+2\Delta^{2})+4\gamma^{2}(\omega_{23}-2\Delta)^{2}
+\Omega^{4}}\,,
\end{equation}
\begin{equation}
\eta(\Delta )=-\frac{2\Delta\Omega(\omega_{23}-\Delta)(\omega_{23} - 2\Delta)} 
{4\Delta^{2}(\omega_{23}-\Delta)^{2} + 2\Omega^{2}(\omega_{23}^{2}-
2\omega_{23}\Delta+2\Delta^{2})+4\gamma^{2}(\omega_{23}-2\Delta)^{2}
+\Omega^{4}}\,. \label{eq:dispersion}
\end{equation}
It is clear from (\ref{eq:absoption}) that when the laser is tuned midway 
between the two levels $\ket{2}$ and $\ket{3}$ (i.e., $\Delta = \omega_{23}/2$) 
the steady-state absorption becomes identically zero due to the destructive 
interference between the amplitudes of the oscillating dipoles of the two 
transitions. On the other hand, if the dipoles are orthogonal to each other 
($\vec d_{21} \cdot \vec d_{31}=0$) we do not observe cancellation of the 
absorption at $\Delta = \omega_{23}/2$. Similarly, equation 
(\ref{eq:dispersion}) shows that the zeros of the dispersion spectrum are 
located at $\Delta=0$, $\omega_{23}/2$, and $\omega_{23}$, respectively. 
Figure~\ref{fig:Fig02} shows the absorption and dispersion spectra as a function 
of the laser detuning with the conditions $|\vec d_{21}| = |\vec d_{31}|$, 
$\gamma_{1} = \gamma_{2} = \gamma= 0.5$, $\Omega_{1} = \Omega_{2} = 0.5$, and 
$\omega_{23} = 1$.

The above model is easier to visualize in the dressed-state representation. For 
simplicity, we keep the condition $\Delta = \omega_{23}/2$, and the magnitudes 
of both dipole moments identical (i.e., $\Omega_{1}=\Omega_{1}=\Omega$ and 
$\gamma_1 = \gamma_2 =\gamma$). The eigenvalues of the interaction Hamiltonian 
are $Z_a=-\Omega_{\mathit{eff}}/2$, $Z_b=0$, and $Z_c=-\Omega_{\mathit{eff}}/2$ 
(where $\Omega_{\mathit{eff}}=\sqrt{(\omega_{23})^2+8(\Omega)^2}$). We further 
assume that $\Omega_{\mathit{eff}}$ is greater than all relaxation rates. The 
steady state population of the dressed states are given by~\cite{Zhou:1997.56}
\begin{eqnarray}
\rho_{aa} &=& (\Gamma_a -\Gamma_{c})/2\Gamma_{a} = \rho_{cc}, \nonumber\\
\rho_{bb} &=& \Gamma_{c}/\Gamma_{a},
\end{eqnarray}
with $\Gamma_a $ and $\Gamma_{c}$ representing the decay constants related to 
the dressed states, and expressed as
\begin{eqnarray}
\Gamma_{a} &=& (\gamma +\gamma_{12})y^2+(\gamma 
-\gamma_{12})(3y^{4}-4y^2+2), \nonumber\\
\Gamma_{c} &=& \half\left[(\gamma +\gamma_{12})y^2+(\gamma 
-\gamma_{12})y^{4})\right], \nonumber \\
y &=& \omega_{23} / \Omega_{\mathit{eff}}.
\end{eqnarray}
If maximum quantum interference is present in the system, then $\gamma_{12} = 
\gamma$, $\rho_{aa}=\rho_{cc}=0$, $\rho_{bb}=1$, the population is entirely 
trapped in the dressed state $\ket{b}$, and there is zero absorption at $\Delta 
= \omega_{23}/2$. Thus the destructive quantum interference is responsible for 
this zero absorption. On the other hand, if $\gamma_{12}=0$ then the absorption 
is almost near maximum depending upon the value of $\Omega $ with respect to 
$\omega _{23} $. However, if $\gamma_{12}$ is slightly lower than its maximum 
value $\gamma $, then destructive quantum interference will not be complete but 
still a considerable amount of population will be trapped in the dressed state 
$\ket{b}$ and some population will be available in the dressed states $\ket{a}$ 
and $\ket{b}$. Consequently, some decay of population from the dressed state 
$\ket{b}$ takes place with decay constant equal to $\gamma y^{2}$. If $y \ll 
1$, then the splitting $\omega_{23}$ of the upper two levels is much less than 
the effective Rabi frequency $\Omega_{\mathit{eff}}$, and the decay constant of 
the dressed state $\ket{b}$ is much smaller than $\gamma$, i.e., $\Gamma_b \ll 
\gamma$. Such small decay constant is responsible of producing narrow Lorentzian 
peaks in the fluorescence spectrum~\cite{Zhou:1997.56}. At this stage, we would 
like to emphasize that for observing population trapping in the dressed state we 
need not have two Rabi frequencies to be equal. As long as the dipole moments of 
two transitions are parallel and $\Delta = \omega_{23}/(1+a^{2})$, with $a$ a 
real number, the population trapping would take place irrespectively of the 
actual value of $a$, $\omega_{23} $ and $\Omega_{\mathit{eff}}$.

\begin{figure}[ht]
  \centering
  \includegraphics{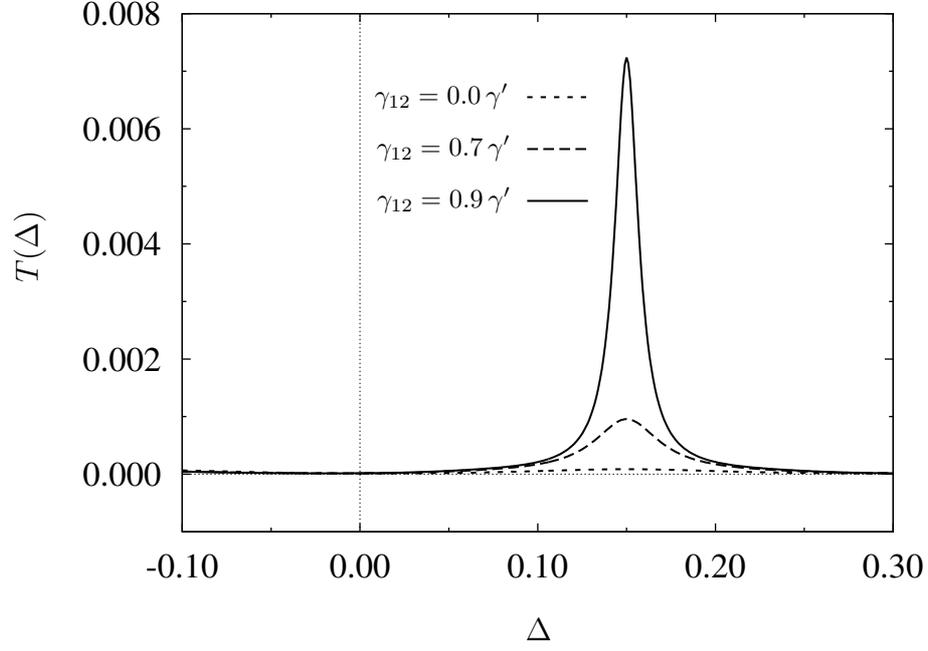}
  \caption{Transmitted light intensity from the optical cavity versus atomic
  detuning $\Delta$ for different values of the induced quantum interference
  $\gamma_{12}$, with $\gamma' = \sqrt{\gamma_1 \gamma_2}$. The spontaneous 
  decay constants are $\gamma_1 = \gamma_2 = 0.05$, the Rabi frequencies 
  $\Omega_1 = \Omega_2 = 0.1$, and the level splitting $\omega_{23} = 0.3$.
  Maximum transmission occurs at $\Delta = 0.15$.}
  \label{fig:Fig03}
\end{figure}

In the discussion above, we keep the atomic system in a vapor cell of unity 
length in an optical cavity of length $\ell$. We can separate the susceptibility 
$\chi$ of the medium in its imaginary and real parts, $A$ and $\eta$, as 
mentioned above in (\ref{eq:absoption}) and (\ref{eq:dispersion}), respectively. 
The absorption coefficient of the medium is related with the imaginary part: 
$\alpha = (n_{0}\omega_{L}/c)A$, in which $n_{0}$ is the refractive index of 
background. The resonant frequency of the cavity is pulled due to the dispersion 
caused by the intracavity medium, and in accordance with the relation
\begin{equation}
\omega_{r} = \frac{1}{1+\zeta}\omega_{C}+\frac{\zeta}{1+\zeta}\bar\omega,
\end{equation}
where $\omega_C$ is the resonant frequency of the empty cavity and $\bar\omega$ 
is the average of the two atomic transition 
frequencies~\cite{Lukin:1998.23,Wang:2000.25,Goorskey:2002.49}. The parameter 
$\zeta = \omega_{r}(s/2\ell)({\partial\chi}'/{\partial\omega_{L}})$ is the 
change in dispersion with respect to the laser frequency. It is easy to show 
that due to the presence of a dispersive medium in the cavity, there is a change 
in the linewidth of the cavity resonance over its empty cavity linewidth. The 
ratio of these two linewidths is given by
\begin{equation}
\frac{\gamma_{m}}{\gamma_{e}}=\frac{1-R\kappa}
{\sqrt{\kappa}(1-R)}\frac{1}{1+\zeta},
\end{equation}
in which $\gamma_{e}$ is the linewidth of the empty cavity, $\gamma_{m}$ is the 
linewidth of the cavity with the medium, $R$ is the reflectivity of both 
mirrors, and $\kappa = \exp(-\alpha s)$ is related to the absorption in a single 
pass. Figure 7 of Ref~\cite{Goorskey:2002.49} clearly depicts the experimental 
demonstration of line narrowing in cavity transmission profile due to the 
dispersive medium present in the cavity (Fig.~\ref{fig:Fig02} of this manuscript 
shows the dispersion). Note that, if we consider the two-level atomic medium in 
the cavity and say the $\omega_{r}$ is near to the atomic transition frequency, 
the dispersion becomes larger causing a narrowing of the linewidth, but at the 
same time the absorption also becomes larger cancelling the narrowing effect. In 
the case of the atomic model under consideration, if coherence (SGC) is present 
in the system (parametric condition of Fig.~\ref{fig:Fig02}) then the cavity 
transmission spectrum shows line narrowing as depicted in Fig.~\ref{fig:Fig03} 
for the central peak (see Ref~\cite{Zhou:1997.56}).

\begin{figure}
  \centering
  \includegraphics{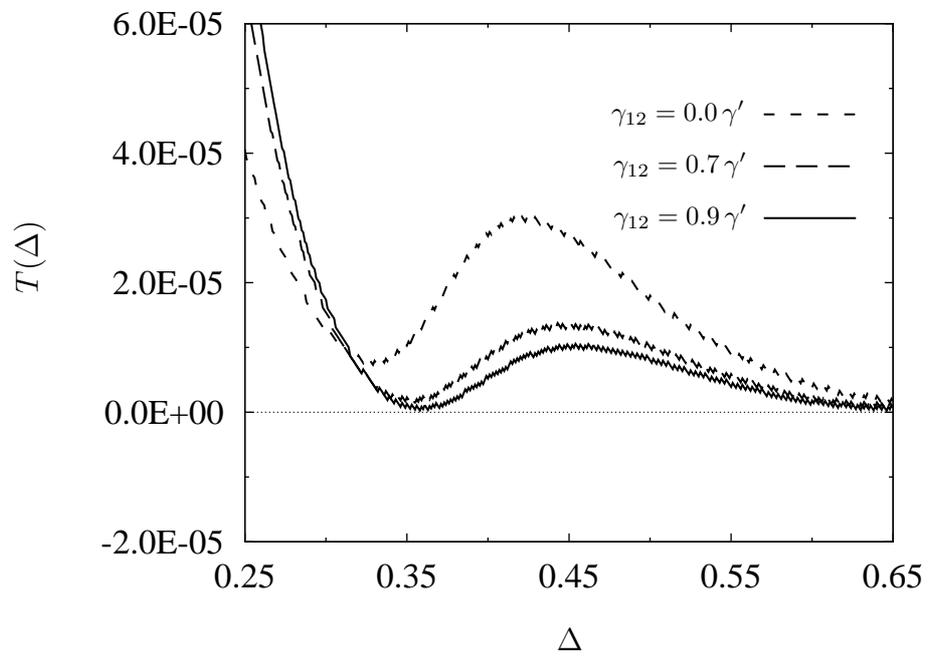}
  \caption{Peak of the vacuum Rabi-splitting spectrum (VRS) for the same system 
  of Fig.~\ref{fig:Fig03}, and including SGC.}
  \label{fig:Fig04}
\end{figure}

Having established the reduction of cavity's and dressed state's decay rates, we 
are certainly in a situation that allows to carry out cavity QED experiments in 
the optical regime without the need of decreasing the cavity mode volume or 
requiring high finesse of the optical cavity. One such example of strong 
coupling cavity QED experiments in optical regime is the observation of vacuum 
Rabi-splitting (VRS) 
spectrum~\cite{Zhu:1990.64,Thompson:1992.68,Agarwal:1984.53, Berman:1994}. In 
Fig.~\ref{fig:Fig04}, we showed one peak of the VRS spectrum (the other peak is 
symmetrically located on the other side) for the V-system (including SGC) under 
consideration within the formalism of linear absorption-dispersion theory and 
interference of multiple beams~\cite{Zhu:1990.64,Thompson:1992.68}. We obtained 
the location of the VRS peaks using the the dressed state picture. The 
eigenvalues and the 
eigenstates~\cite{Cardimona:1989.22,Zhou:1996.77,Braunstein:2001.64} for the 
system are
\begin{eqnarray}
\left[0, \frac{2g\sqrt{n+1}}{\Omega_N}\ket{2,n}-
\frac{2g\sqrt{n+1}}{\Omega_N}\ket{3,n}+
\frac{\omega_{23}}{\Omega_N}\ket{1,n+1}\right], \nonumber\\
\left[\frac{\Omega_{N}}{2}, 
\half\left(1-\frac{\omega_{23}}{\Omega_N}\right)\ket{2,n}+
\half\left(1+\frac{\omega_{23}}{\Omega_N}\right)\ket{3,n}+
\frac{2g\sqrt{n+1}}{\Omega_N}\ket{1,n+1}\right], \nonumber\\
\left[-\frac{\Omega_{N}}{2}, 
-\half\left(1-\frac{\omega_{23}}{\Omega_N}\right)\ket{2,n}-
\half\left(1+\frac{\omega_{23}}{\Omega_N}\right)\ket{3,n}+
\frac{2g\sqrt{n+1}}{\Omega_N}\ket{1,n+1}\right], \nonumber
\end{eqnarray}
where $\Omega_{N}=\half[\omega_{23}+8g^{2}(n+1)]^{1/2}$. The VRS spectrum is 
the transition from ground state to the first manifold of the dressed 
state~\cite{Agarwal:1984.53}. Clearly, the peaks are located at $\omega_{23}/2$, 
$\omega_{23}/2\pm\sqrt{2}g$.

To achieve SGC in this system, we need to have parallel dipole moments for the 
two transitions, which is difficult to achieve practically. However, in a recent 
proposal (see Ref~\cite{Ficek:2004.69}), it is claimed that there is no 
need to have parallel dipole moments to achieve strong quantum interference in 
such system. One can work with the perpendicular dipole moments but it requires 
cw field to drive the transition between the upper atomic states. This system 
exhibits the same features shown by that with parallel dipole moments. We expect 
that this work could be applicable for cavity cooling experiments of single 
atoms~\cite{Maunz:2004.428}.

\section{\label{sec:conclusion}Conclusions}
In this work, we proposed an alternative scheme to carry out a strong coupling, 
cavity quantum electrodynamics experiments in the optical regime using 
spontaneously generated coherence and enhanced dispersion. To do this, we 
considered a V-type, three-level atomic system interacting with a single mode 
field inside an optical cavity. Then, we showed by a sample calculation that the 
spontaneously generated coherence and enhanced dispersion in this system 
considerably reduced the radiative and cavity decay rates when the laser is 
tuned midway between the two excited levels $\ket{2}$ and $\ket{3}$ (i.e., 
$\Delta = \omega_{23}/2$), maximizing the quantum interference in the system. 
However, there was no restriction in the selection of $\Delta$.

To elaborate our proposal quantitatively, we compared
our results with the 
following fundamental rates of strong coupling regime obtained by Thompson et 
al.~\cite{Thompson:1992.68}: $(g, \gamma_{\perp}, \gamma_{cav}) = [2\pi (3.2, 
2.5, 0.9)]\units{MHz}$ (here, $\gamma_{\perp} = \gamma_a / 2$). The $g$ can be 
calculated from its definition $g = (\mu^2 \omega_0 / 2\hbar\epsilon_0 
V)^{1/2}$. Clearly, $g$ depends on the cavity-mode volume $V = \pi w_0^{2} \ell 
/ 4$, where $w_0$ is the mode waist and $\ell$ is the length of the cavity. In 
our proposal, the atomic decay constant decreased by a factor between 10 and 100 
due to SGC~\cite{Tian:2012.285,Song:2016.015001,Zhou:1997.56}; the cavity decay 
constant also decreased by a factor of 
14~\cite{Lukin:1998.23,Wang:2000.25,Goorskey:2002.49} due to enhanced dispersion 
and requiring a lower finesse cavity; and the critical photon number and atom 
number remain more or less the same. Therefore, we can have $g$ lowered by a 
factor of 10. This means one can work with a higher mode volume this time. 
Consequently, we may have a cavity length and a mode waist 20 and 2.2 times 
larger than those used in that experiment, respectively. Perhaps, this will 
allow the use of larger cavities with lower finesses.  cavity. This observation 
offers a new perspective in cavity quantum electrodynamics experiments as it 
shows that the use of expensive miniaturized cavity with high finesse may not 
longer be required.

\section{\label{sec:ack} Acknowledgements}
We thank M. Xiao for his suggestions and helpful discussions.

\section*{References}

\newpage
\clearpage

\clearpage

\clearpage

\clearpage

\end{document}